\documentclass[11pt]{article}

\usepackage[preprint]{acl}

\usepackage{times}
\usepackage{latexsym}

\usepackage[T1]{fontenc}
\usepackage[utf8]{inputenc}

\usepackage{microtype}

\usepackage{inconsolata}

\usepackage{graphicx}
\usepackage{booktabs}
\usepackage{multirow}
\usepackage{amsmath}
\usepackage{amssymb}
\usepackage{cleveref}
\usepackage{fvextra}
\usepackage{array}
\usepackage{tabularx}

\title{Be Careful Who You Trust: Coordination Dynamics under Corrupted Communication in LLM Multi-Agent Games}

\author{
  Xuanyi Liu \quad
  Niall Dalton \quad
  Hairi Amin\thanks{Equal contribution.} \quad
  Xiyuan Yin\footnotemark[1] \quad
  Lydia Lim \\
  Department of Computer Science, University College London \\
  \small{
    \textbf{Correspondence:}
    \href{mailto:niall.dalton.25@ucl.ac.uk}{niall.dalton.25@ucl.ac.uk}
  }
}

\begin{document}
\maketitle
\begin{abstract}
Large language models are increasingly used as interacting agents, but it remains unclear how robust their coordination is when public communication is unreliable. We study this question in iterated \(N\)-player Stag Hunt games played by homogeneous LLM groups under controlled programmatic action inversion, which changes both the public transcript and the actions used for execution. Across an experimental grid spanning group sizes, coordination thresholds, corruption levels, and seven LLMs, we observe three main patterns. First, honest agents' pre-flip Stag choices decline as corruption increases, but the sharp fall in public success is primarily mechanical. In the focal \(N=5,M=3\) setting, pre-flip success remains 78\% at 80\% corruption, while public success falls to 12\%. Second, honest choices are associated with the public history available at decision time, particularly under high corruption. Third, three threshold-style public-report benchmarks yield similar action-match rates to the LLM agents, showing substantial descriptive agreement between LLM decisions and these benchmarks. Overall, our results show that original choices, public actions, and executed outcomes must be separated when evaluating multi-agent robustness, as corrupted communication can severely and predictably degrade mutually beneficial cooperation.
\end{abstract}

\section{Introduction}

Coordination under uncertainty is fundamental to multi-agent systems because collective outcomes depend on both agents' decisions and the information available to them. Classical game theory shows that changes in information or incentives can shift coordination between payoff-dominant and risk-dominant equilibria \citep{Harsanyi1988,Kandori1993,CarlssonHans1993GGaE}. Whether these insights extend to large language model (LLM)-based agents remains unclear, as their decisions arise through language-conditioned inference rather than an explicitly implemented game-theoretic policy \citep{AkataElif2023Prgw}. Recent work shows sensitivity to interaction structure and social context, alongside misleading outputs and error propagation in multi-agent systems \citep{chen2023multiagentconsensus,HaoJianing2026GLoL,shapira2026agents}.

We study an iterated \(N\)-player Stag Hunt played by homogeneous groups drawn from seven LLMs. Agents act sequentially and communicate through structured public reports. For designated corrupted agents, the original action is retained for analysis, while the action shown in the public transcript and used for execution is programmatically inverted before later agents observe it.\footnote{We use \emph{corrupted agent} rather than \emph{adversarial agent}, since the action is transformed post hoc by the simulator rather than reflecting adversarial intent by the model itself.} This separates original pre-flip choices, public post-flip actions, pre-flip outcomes calculated from original choices, and executed public outcomes. Pre-flip success is not a no-corruption counterfactual because later original choices may already have responded to corrupted reports; it isolates only the final mechanical effect of action inversion.

Concretely, suppose Agent A originally chooses STAG. The simulator records STAG as A's pre-flip action for analysis. If A has been designated a corrupted agent for this run, the simulator overwrites the public transcript entry with HARE and regenerates a justification consistent with HARE; Agent B, speaking next, therefore observes only the public HARE report, with no indication that a transformation occurred. At the end of the round, both A's original STAG choice and the public HARE report are retained, so that success and payoffs can be computed separately from each action layer.

Our contributions are threefold. First, we separate changes in honest agents' original choices from the direct public-action channel. Honest Stag choices decline gradually, whereas public success falls much more sharply. Second, we show that honest actions are sensitive to visible transcript information: under high corruption, public-history benchmarks match honest choices more closely than hidden-original-history benchmarks, and sequential-response models show a strong association between prior public Stag share and subsequent choices. We note that these analyses are observational and do not identify causal mediation. Third, we compare observed actions with within-round, carryover, and outcome-alignment-weighted public-report benchmarks and quantify the welfare gap between realised payoffs and truthful implementation of the same recorded choices. Together, these results indicate that a substantial part of the apparent coordination collapse under corruption is mechanical rather than behavioural, underscoring the need to separate original choices, public actions, and executed outcomes when evaluating multi-agent robustness more broadly.

\section{Related Work}
\label{sec:related}

Equilibrium selection in Stag Hunt contrasts a payoff-dominant equilibrium with a risk-dominant equilibrium \citep{Harsanyi1988}. This distinction extends naturally to \(N\)-player and threshold variants, where efficient coordination requires a critical mass of cooperators and may exhibit sharp transitions under particular theoretical assumptions \citep{Pacheco2008}. More broadly, Stag Hunt has been used to formalise ideas about social contracts and convention formation \citep{Skyrms2003}. Prior work on evolutionary and learning dynamics shows that persistent small shocks can select the risk-dominant equilibrium in the long run \citep{Kandori1993}, while behavioural studies of repeated coordination highlight tipping and path dependence between equilibria under noise \citep{mas2016behavioral}. Global games similarly show how small informational imperfections can induce threshold strategies and sharpen equilibrium predictions \citep{CarlssonHans1993GGaE,FrankelDavidM.2003Esig}. Empirically, belief-based predictors can capture behaviour in coordination environments \citep{HEINEMANN2024632}. These theories provide useful analytical reference points, but do not establish that LLM agents implement the corresponding belief dynamics.

Recent work has evaluated LLMs as game-playing agents in repeated games, showing that they can exhibit non-trivial strategic behaviour while remaining sensitive to contextual framing \citep{AkataElif2023Prgw,lore2024strategic}. Moving from single-agent game play to multi-agent interaction, prior work shows that convergence and consensus depend on group size and interaction structure \citep{chen2023multiagentconsensus}. Together, these results motivate our question of how stable LLM coordination remains when the public information available to agents is systematically corrupted.

Robustness work studies how adversarial or Byzantine agents affect decentralised coordination and multi-agent collaboration. This includes Byzantine-tolerant coordination and robust aggregation methods \citep{JoYongrae2025BDCo}, attack-focused analyses \citep{amayuelas2024multiagentattack}, and benchmarks that catalogue adversarial risks and failure modes \citep{kavathekar2025tamas}. Other studies show that LLM responses are sensitive to framing, presentation order, and strategically sequenced evidence \citep{moore2025askwhaiprobingbeliefformation,hu2026lyingtruthsopenchannelmultiagent}. Evidence from social deduction games further shows substantial heterogeneity across models in deception and deception-detection performance \citep{curvo2025traitorsdeceptiontrustmultiagent,yoo2024finding}.

Our work relates to these threads but differs in emphasis. Rather than treating a sharp decline in executed coordination as direct evidence of an emergent behavioural tipping point, we separate agents' original choices, the public actions visible to later agents, and the threshold outcomes computed from each action layer. We then compare honest-agent choices with the public information actually available at decision time and with analyst-only reconstructions of the hidden original history. This framing focuses on communication and execution robustness and observable response patterns rather than latent belief recovery.

\section{Methods}
\label{sec:methods}

Under incomplete information, we construct a mean-field analytical benchmark inspired by threshold models of games with incomplete information \cite{Harsanyi1968}. In our setting, uncertainty concerns other agents' actions rather than payoff dominance.

Focusing on the iterated $N$-player Stag Hunt, at each round $t$, we associate each agent $i$ with an analytical estimate of the probability that other agents publicly report Stag. Cooperation yields the highest payoff if at least $M$ players report Stag. We denote this public-report estimate as

\[
q_{i}^{t} = P(\text{public Stag report}).
\]

Here, $q_i^t$ is constructed retrospectively from the public reports available to agent $i$ at round $t$. It is used as an analytical benchmark and should not be interpreted as a direct measurement of the agent's latent internal belief.

This estimate uses a mean-field threshold approximation in which the number of other agents publicly reporting Stag, denoted $K$, is modelled as

\[
K \sim \text{Binomial}(N - 1, q_{i}^{t}).
\]

This binomial formulation is a mean-field analytical approximation. It does not imply conditional independence among agents and does not separately condition on already observed reports and remaining speakers; it should therefore not be interpreted as an exact sequential best response.

\subsection{Public-Report Benchmarks From Structured Reports}

Agents do not observe others' original actions directly, but receive structured messages containing a binary public action $\tilde r_{j}^{t} \in \{\text{Stag}, \text{Hare}\}$. A subset of agents is designated as corrupted, and their public action is programmatically transformed after the LLM produces its original response. Let $N$ denote the total number of agents and $F$ the number of designated corrupted agents. The proportion of agents whose reports are not transformed is

\[
\alpha = \frac{N - F}{N}.
\]

This quantity characterises the experimental corruption level. However, agents are not told $F$, and $\alpha$ is not used to correct the public-report estimate in the implemented benchmarks. At round $t$, agent $i$ observes $K_{i}^{t}$ public Stag reports among the $n_i^t$ agents whose reports are available at the time of decision. Further, let \(y_j^t \in \{0,1\}\) denote agent \(j\)'s public post-flip report in round \(t\), where \(y_j^t = 1\) indicates a public Stag report.

\paragraph{Naive aggregation.}
A natural benchmark is to estimate the public Stag-report rate by the empirical share of observed Stag reports:

\begin{equation}
\hat q_i^t =
\frac{K_i^t}{n_i^t},
\qquad n_i^t > 0.
\label{eq:naive_update}
\end{equation}

This rule uses only within-round information and assigns equal weight to all observed reports. When $n_i^t=0$, no public-report estimate is defined, and the corresponding turn is excluded from the naive benchmark evaluation.

\paragraph{Carryover updating.}
To allow information from the public reporting environment to persist across rounds, we introduce a prior anchored in the previous round's posterior:

\begin{equation}
q_{\mathrm{prior}}^{t}
=
\lambda \hat q_{\mathrm{post}}^{t-1}
+
(1-\lambda)q_0,
\label{eq:carryover_prior}
\end{equation}

where \(q_0 \in [0,1]\) is a baseline prior and \(\lambda \in [0,1]\) governs the strength of inter-round carryover. The within-round estimate is then formed by pseudo-count pooling:

\begin{equation}
\hat q_i^t
=
\frac{\tau q_{\mathrm{prior}}^t + K_i^t}
{\tau + n_i^t},
\label{eq:carryover_update}
\end{equation}

where \(\tau > 0\) is the prior strength. This specification shrinks the current-round estimate towards the carried-over prior when few reports are observed.

\paragraph{Per-agent outcome-alignment weighting.}
A richer alternative weights speakers according to their previous public report--outcome alignment. Let \(\rho_j^t \in [0,1]\) denote the analytical outcome-alignment weight assigned to agent \(j\) in round \(t\). The weighted estimate is

\begin{equation}
\hat q_i^t
=
\frac{
\tau q_{\mathrm{prior}}^t
+
\sum_{j \in S_i^t} \rho_j^t y_j^t
}{
\tau
+
\sum_{j \in S_i^t} \rho_j^t
}.
\label{eq:trust_update}
\end{equation}

The weights are updated retrospectively from previous public-report alignment with the realised public outcome; the full update equations are given in Appendix~\ref{app:trust-details}.

Taken together, these specifications provide three descriptive public-report benchmarks: purely contemporaneous aggregation, round-to-round persistence in the aggregate reporting environment, and speaker-specific outcome-alignment weighting.

\subsection{Mean-Field Payoff Threshold Benchmark}

Given the analytical estimate $q_i^t$, the probability of successful coordination is

\[
P(\text{success}) = P(K \ge M - 1).
\]

The expected payoff from choosing Stag is then

\[
\mathbb{E}[U_S]
=
R_S \cdot P(K \ge M - 1)
+
R_F \cdot P(K < M - 1),
\]

where $R_S$ is the payoff if the threshold is met and $R_F$ is the payoff otherwise. Let the certain payoff from Hare be $H$.

The benchmark condition for choosing Stag is

\[
\mathbb{E}[U_S] \ge H.
\]

Let

\[
S
=
\sum_{k=M-1}^{N-1}
\binom{N-1}{k}
(q_i^t)^k
(1-q_i^t)^{N-1-k},
\]

which is the probability that at least \(M-1\) of the other \(N-1\) players publicly report Stag. The condition can then be written as

\[
R_S S + R_F(1-S) \ge H.
\]

This inequality implicitly defines a threshold
$q^* = f(N,M,R_S,R_F,H)$,
such that the benchmark action has a cutoff form:

\[
\text{Choose Stag if and only if } q_i^t \ge q^*.
\]

Comparative statics follow directly from this formulation \cite{Pacheco2008}.

\subsection{Effect of Observed Cooperation Levels}

Lower visible cooperation produces a lower estimate $\hat q_i^t$ and may therefore change the action implied by the cutoff benchmark. This provides a descriptive reference for observed LLM actions, but does not establish that agents internally maintain or update the same scalar belief. Additional derivations and the treatment of turns without prior reports are given in Appendix~\ref{app:additional-derivations}.

\section{Experiments}
\begin{figure*}[t]
    \centering
    \includegraphics[width=0.9\linewidth]
    {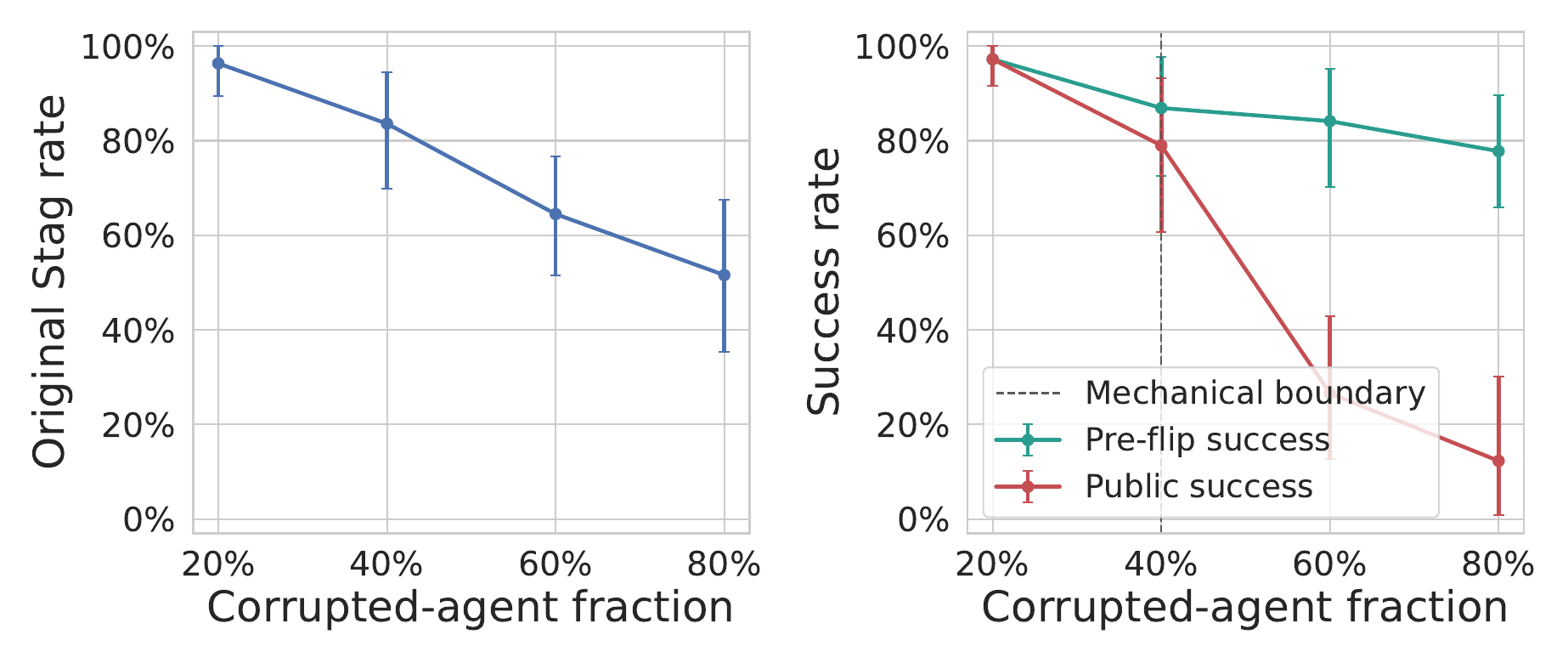}
    \caption{
    Coordination outcomes in the focal setting with $N=5$ and $M=3$.
    The figure separates honest agents' original Stag choices, pre-flip
    success calculated from all original pre-flip actions, and public success
    calculated from the post-flip actions used by the simulation. Both
    success measures are computed from the same corruption-exposed
    trajectories: pre-flip success is not a separate clean or no-corruption
    baseline, since later pre-flip actions may already reflect exposure to
    earlier corrupted reports. The dashed
    line marks the boundary beyond which corruption exceeds the fraction needed for success. Pre-flip
    success does not show the same sharp decline as public success, indicating
    that the apparent threshold in the public outcome is driven substantially
    by deterministic action inversion.
    }
    \label{fig:success-vs-liar-highlight}
\end{figure*}

\subsection{Game Formulation}

We study the iterated $N$-player Stag Hunt under incomplete information. At each round $t$, every agent $i \in \{1, \ldots, N\}$ selects an original action
$r_i^t \in \{\text{Stag}, \text{Hare}\}$. Agents act sequentially and may
condition their decisions on the public reports already available in the
transcript. Coordination succeeds if at least $M$ public actions are Stag, where
$M \in \{1, \ldots, N\}$ is a fixed threshold within each trial. The payoff structure satisfies

\[
R_S > H_F \geq H_S > R_F,
\]

where $R_S$ is the payoff to a player choosing Stag when the threshold is met, $H_F$ is the payoff to a player choosing Hare when the threshold is not met, $H_S$ is the payoff to a player choosing Hare when the threshold is met, and $R_F$ is the payoff to a player choosing Stag when the threshold is not met. This ordering corresponds to the standard two-player Stag Hunt inequalities $u(\text{Stag}, \text{Stag}) > u(\text{Hare}, \text{Hare}) \geq u(\text{Hare}, \text{Stag}) > u(\text{Stag}, \text{Hare})$, where $u(a,a')$ denotes a player's payoff when choosing action $a$ while a co-player chooses action $a'$. We extend this ordering to $N$ players via the threshold formulation \cite{Pacheco2008}. For $M\geq2$, the game admits universal Stag (payoff-dominant) and universal Hare (risk-dominant) as pure Nash equilibria; for $M=1$, universal Hare is not an equilibrium. Our experiments vary $N$, $M$, the number of corrupted agents $F$, and
the number of rounds $T$, while holding the payoff tuple
$(R_S,H_S,R_F,H_F)$ fixed.

\subsection{Agent Design}

Each player is a large language model (LLM) agent whose only inputs are a fixed system prompt and the accumulating public transcript. The system prompt (available in Appendix~\ref{app:prompts})  specifies the agent's name, the game parameters $N$ and $M$, and the full payoff matrix, but provides no private information about other agents. Agents are not informed of the number or identities of corrupted agents. Agents can only communicate through the public transcript.

Within each round, agents speak sequentially in a fixed order. When it is agent $i$'s turn, it receives all messages produced so far in that round. This includes the round-start announcement from the Game Master and the structured reports of all agents who have already spoken. Agent $i$ then produces a structured output containing a binary action $r_i^t \in \{\text{Stag}, \text{Hare}\}$, a scalar confidence $c_i^t \in [0, 1]$, and a one-sentence text justification. We refer to $r_i^t$, before any programmatic transformation, as the
agent's original or pre-flip action. Across rounds, the full transcript is retained and extended, such that agents in round $t$ can observe every report and justification from all prior rounds.

\subsection{Corruption Mechanism}

A subset of $F$ agents is designated as corrupted at the start of each simulation, with identities drawn uniformly at random from the $N$ agents and held fixed across all rounds. Since corrupted-agent identity is undisclosed, corrupted agents are indistinguishable from honest agents from the perspective of the public transcript. The corruption mechanism operates post hoc and programmatically, entirely outside the LLM's initial decision process. After a designated corrupted agent produces its original decision $(r_i^t, c_i^t, \text{just}_i^t)$, the simulation inverts the action shown in the public transcript:

\[
\tilde{r}_i^t = \begin{cases}
\text{Hare} & \text{if } r_i^t = \text{Stag} \\
\text{Stag} & \text{if } r_i^t = \text{Hare}
\end{cases}
\]

As the flipped action differs from the original, a separate LLM call using the same model as the corrupted agent generates a new justification that is plausible for the flipped action. The original confidence score is retained. The public transcript entry is then patched to replace the original output with $(\tilde{r}_i^t, c_i^t, \widetilde{just}_i^t)$ before the next agent in the speaking order receives any messages. Subsequent agents thus receive the transformed report without an explicit indicator of corrupted status.

Consequently, round outcomes and payoffs are determined by the publicly reported actions $\tilde{r}_i^t$ rather than the agents' original actions $r_i^t$. Public Stag coordination succeeds if and only if

\[
S_{\mathrm{pub}}^t
=
\mathbb{I}
\left[
\sum_i
\mathbb{I}[\tilde r_i^t=\text{Stag}]
\geq M
\right].
\]

For analysis, we also calculate pre-flip success by applying the same coordination threshold to the recorded original actions (we retain the notation $S_{\mathrm{int}}^t$ for this quantity):

\[
S_{\mathrm{int}}^t
=
\mathbb{I}
\left[
\sum_i
\mathbb{I}[r_i^t=\text{Stag}]
\geq M
\right].
\]

Pre-flip success is not a no-corruption counterfactual, because later agents' original actions may already have responded to earlier corrupted public reports. Instead, it isolates whether the final programmatic action inversion changes the threshold outcome.

The joint transcript-and-execution intervention therefore has two distinct effects: it changes the information available to subsequent agents and directly changes the public actions used to calculate coordination success and realised payoffs. We retain both original and public actions so that these effects can be reported separately.

\subsection{Experimental Protocol}

\paragraph{Parameter grid.}
The primary analysis uses fixed speaking order, deterministic report inversion, and homogeneous model groups. The realised support primarily includes $N\in\{2,3,5\}$, $T\in\{1,2,4\}$, and $M\in\{1,\ldots,N\}$. For $N=2$, the observed corrupted-agent counts are $F\in\{0,1\}$; for $N=3$, $F\in\{1,2\}$; and for $N=5$, $F\in\{1,2,3,4\}$. The grid is unbalanced, and the $N=3$ and $N=5$ settings do not contain matched fully honest $F=0$ conditions. Comparisons across corrupted-agent fractions are therefore descriptive rather than fully matched causal contrasts.

Payoffs are fixed across all conditions at $(R_S, H_S, R_F, H_F) = (4, 2, 0, 2)$, which satisfies the Stag Hunt ordering and produces a cooperation threshold $q^*$ that varies with $N$ and $M$ alone. We evaluate seven LLM agents: \texttt{deepseek-v3.2-think}, \texttt{ernie-5.0-thinking-preview}, \texttt{glm-5}, \texttt{gpt-5-mini}, \texttt{gpt-5.2-2025-12-11}, \texttt{kimi-k2.5}, and \texttt{llama-3.1-8b}. Notably, we restrict each trial to a homogeneous model pool; that is, all agents in a trial are of the same model.

\subsection{Evaluation Metrics}

We report metrics separately for original pre-flip actions, public post-flip actions, and outcomes calculated from each action layer.

\paragraph{Honest original Stag rate.}
We report the proportion of honest-agent turns for which the original pre-flip action $r_i^t$ is Stag, separating behavioural changes from the direct transformation applied to corrupted agents.

\paragraph{Pre-flip and public success.}
Pre-flip success applies the coordination threshold to original actions, while public success applies it to public post-flip actions, as defined above. We also distinguish shared success, flip-induced loss, flip-induced rescue, and shared failure from the joint values of $S_{\mathrm{int}}^t$ and $S_{\mathrm{pub}}^t$.

\paragraph{Payoff.}
We report mean realised payoff per honest-agent turn. We also calculate a truthful-implementation payoff by applying the payoff rule to the same recorded original actions and pre-flip outcome.

Definitions of outcome alignment, confidence calibration, downstream report alignment, and consensus entropy are provided in Appendix~\ref{app:additional-metrics}.

\begin{figure*}[t]
    \centering
    \includegraphics[width=0.9\linewidth]{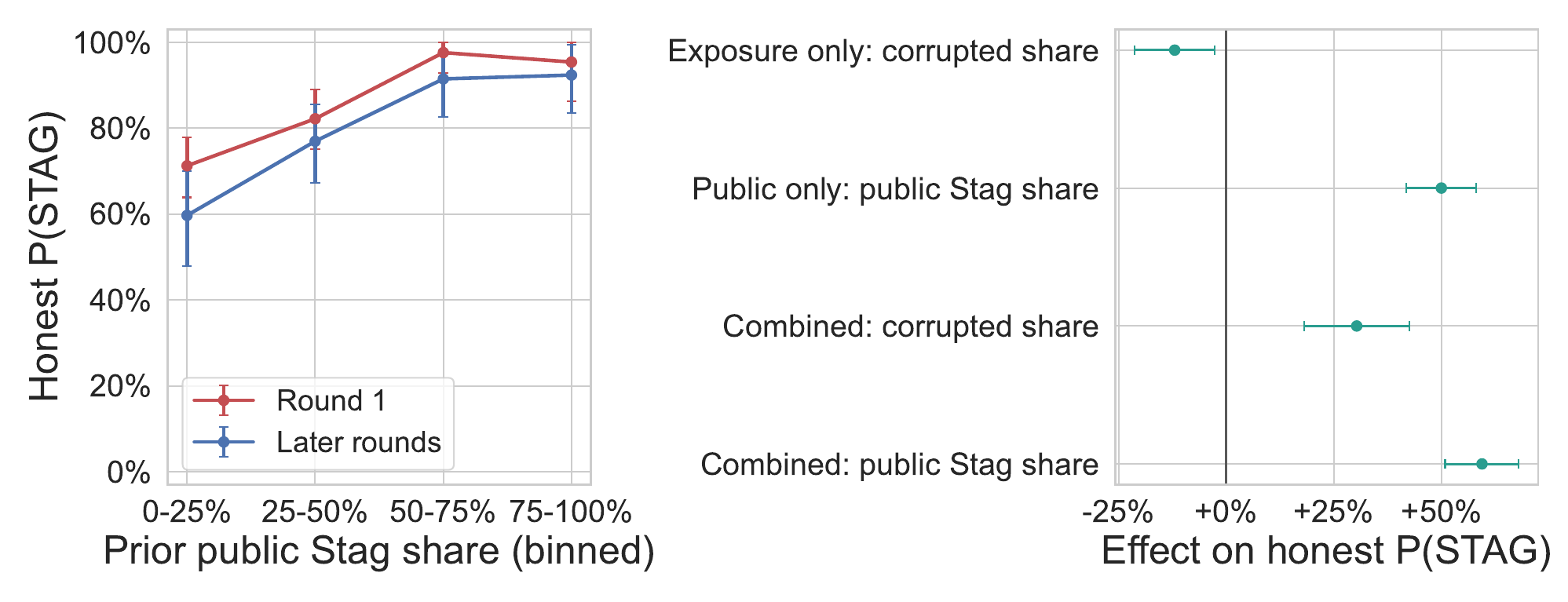}
    \caption{Sequential-response analysis for honest agents' original choices.
    The left panel relates prior public Stag share to the probability of
    originally selecting Stag across prior corruption-exposure levels. The
    right panel reports fixed-effect linear probability estimates with 95\%
    confidence intervals.}
    \label{fig:sequential-exposure}
\end{figure*}

\subsection{Analytical Benchmark Evaluation}

To evaluate descriptive agreement between observed LLM actions and the analytical benchmarks, we consider each honest-agent turn for which at least one prior public report is available. For each of the three rules in \S\ref{sec:methods}, we form the rule-implied estimate $\hat{q}_i^t$ and derive the benchmark action as Stag if $\hat{q}_i^t \geq q^*$ and Hare otherwise. The match rate is then the proportion of eligible turns on which the agent's original pre-flip action agrees with the action implied by the benchmark.

This comparison measures descriptive behavioural alignment. It does not establish that the agent internally computes $\hat{q}_i^t$, represents $q^*$, maintains the corresponding latent belief, or follows a Bayesian update.

For the carryover and outcome-alignment-weighted rules, we fix $\lambda = 0.5$, $q_0 = 0.5$, $\tau = 2$, and initialise outcome-alignment counts at $(a_j^0, b_j^0) = (1, 0)$. All three benchmarks are compared on common-support turns with at least one previously observed public report.

\section{Results and Discussion}

\subsection{Coordination Performance under Corrupted Public Reports}
\label{sec:Coordination}

Figure~\ref{fig:success-vs-liar-highlight} shows that public coordination declines as the corrupted-agent fraction increases, with a sharp drop as corruption moves beyond the all-Stag inversion boundary $1-M/N$ (40\% for $N=5, M=3$). However, pre-flip success does not show the same sharp decline. In the focal setting, pre-flip success decreases from 0.972 at 20\% corruption to 0.778 at 80\%, whereas public success falls from 0.972 to 0.123. The apparent threshold in public success is therefore driven substantially by deterministic action inversion rather than a corresponding behavioural tipping point.

Appendix Figure~\ref{fig:coordination-full} shows how these measures vary across rounds. Under lower corruption, honest original Stag choices, pre-flip success, and public success generally remain high. Under heavier corruption, public success is lower and more variable, while pre-flip success and honest original Stag choices do not exhibit the same uniform collapse.

Overall, honest agents' original Stag choices decline with corruption, while public outcomes deteriorate more sharply because the procedure directly changes the actions used to determine success. Model-level and cross-game results are reported in Appendix Figures~\ref{fig:mechanism-by-model} and~\ref{fig:coordination-full-plot}.

\subsection{Information Dynamics and Sequential Response Patterns}

Appendix Figure~\ref{fig:turn-order} shows that the association between speaking
position and honest agents' original Stag choices varies across corruption
levels. Because speaking order is fixed in the primary condition, these
patterns are descriptive and do not establish that additional information
itself improves performance or amplifies noise.

Appendix Figure~\ref{fig:public-vs-original-history} compares threshold
predictions based on the public reports available at decision time with
predictions based on the corresponding hidden original-action history. In the
focal $N=5,M=3$ setting, hidden-original history matches honest choices more
closely at 20\% and 40\% corruption, whereas public history matches more
closely at 60\% and 80\%. The comparison therefore does not show a universal
advantage for either history, but supports sensitivity to the visible
transcript under heavier corruption.

Figure~\ref{fig:sequential-exposure} provides complementary turn-level
evidence. In the public-history-only specification, a unit increase in prior
public Stag share is associated with a 0.499 increase in the probability of
selecting Stag (95\% CI $[0.418,0.580]$). In the joint specification, the
corresponding coefficient is 0.594 (95\% CI $[0.508,0.679]$). The coefficient
on prior corruption exposure changes from $-0.120$ (95\% CI
$[-0.213,-0.026]$) in the exposure-only specification to 0.303 (95\% CI
$[0.180,0.425]$) after conditioning on public Stag share.

\begin{figure*}[tb]
    \centering
    \includegraphics[width=\linewidth]{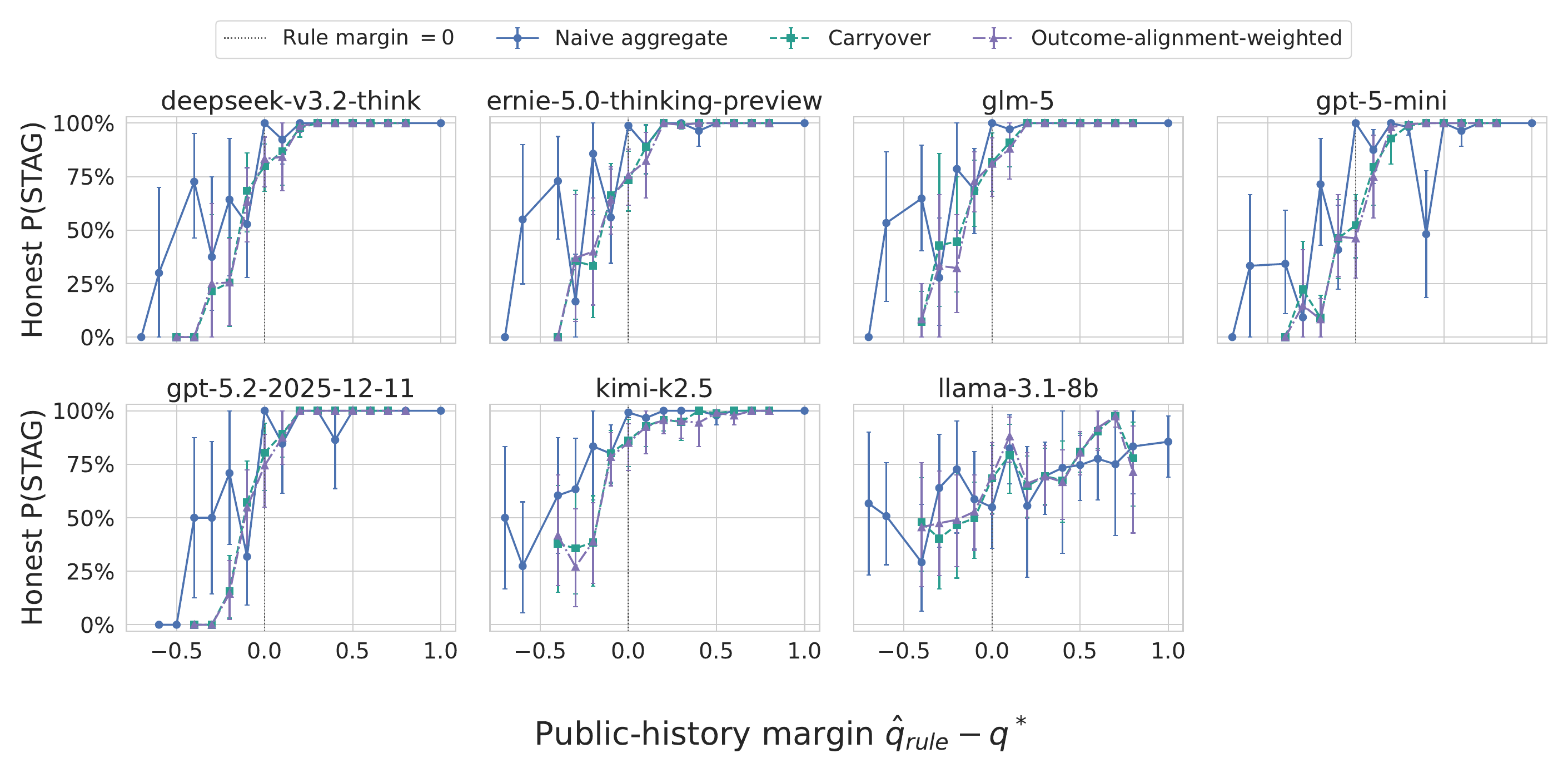}
    \caption{Probability that an honest agent originally selects Stag as a
    function of the public-report benchmark margin
    $\hat{q}_{\mathrm{rule}}-q^*$, shown for the within-round, carryover, and
    outcome-alignment-weighted rules. Here $\hat{q}_{\mathrm{rule}}$ is the
    benchmark's estimated probability that other agents publicly report Stag
    (\S\ref{sec:methods}), and $q^*$ is the payoff-implied cutoff at which
    the expected payoff from Stag equals the safe Hare payoff. The vertical
    zero point marks this action threshold: a positive margin
    ($\hat{q}_{\mathrm{rule}}-q^*>0$) means the benchmark recommends Stag,
    and a negative margin means it recommends Hare.}
    \label{fig:belief-response}
\end{figure*}

\subsection{Alignment with Analytical Threshold Benchmarks}
\label{sec:benchmark-alignment}

For each eligible honest-agent turn, we compute the action implied by each
analytical benchmark via $\hat{q}_{\text{rule}}$ and report the fraction of
turns on which the agent's original pre-flip action matches the
benchmark-implied action (match rate).

\begin{table}[h]
    \centering
    \caption{Overall analytical benchmark match rates on honest-agent turns
    with common support across all three rules.}
    \label{tab:belief-summary}
    \begin{tabular}{lc}
        \toprule
        \textbf{Benchmark Rule} & \textbf{Match Rate} \\
        \midrule
        Naive aggregate          & 0.829 \\
        Carryover                & 0.846 \\
        \textbf{Outcome-alignment}  & \textbf{0.850} \\
        \bottomrule
    \end{tabular}
\end{table}

\paragraph{Benchmark match rates.}
Across 2,852 eligible observations, match rates increase from 0.829
(naive) to 0.846 (carryover) and 0.850 (outcome-alignment-weighted). The carryover
and outcome-alignment-weighted benchmarks therefore provide slightly higher descriptive
agreement than the naive aggregate benchmark, although the differences are
small. Model-level results (Table~\ref{tab:belief-match-rate}) show substantial
heterogeneity. Several models reach match rates of 0.86--0.96, and five of
the seven models perform best under the outcome-alignment-weighted benchmark.
\texttt{Llama-3.1-8b} has lower match rates of approximately 0.62 across
all three rules, indicating model-specific variation rather than an effect
that can be attributed to the benchmark specification alone.

\paragraph{Benchmark-margin analysis.}
Figure~\ref{fig:belief-response} plots $P(\text{Stag})$ against the margin
$\hat{q}_{\text{rule}}-q^*$. For six of the seven models, the probability
of selecting Stag generally increases as the benchmark margin moves above
zero across the three rules. This pattern is consistent with
threshold-sensitive behaviour, showing that most models behave in a manner
broadly consistent with the benchmarks, but does not establish that the
models internally represent said benchmark rules.
\texttt{Llama-3.1-8b} again shows a shallower response and no clear
threshold crossing.

\subsection{Payoff Outcomes and Robustness to Interventions}

As the fraction of corrupted agents increases, average realised payoffs
drop substantially. In the available fully honest settings, agents approach
the theoretical maximum of 4.0, while under high corruption payoffs fall to
roughly 2.0--2.4 (Table~\ref{tab:overall-performance}). Because realised
payoffs are calculated from public post-flip actions, part of this decline
follows directly from deterministic action inversion. The concurrent decline in honest agents' original Stag choices suggests an additional behavioural component whose causal contribution is not identified.

Appendix Figure~\ref{fig:welfare-decomposition} shows that the gap between
realised and truthful-implementation payoff grows under heavier corruption.
At 60\%, realised payoff is 1.702 while truthful-implementation payoff remains
3.218; at 80\%, the corresponding values are 1.444 and 3.032. The widening
gap indicates a substantial mechanical component, alongside the decline in
honest agents' original Stag choices.

\section{Conclusion}

Observed coordination failure under corrupted communication mixes two
distinct phenomena: changes in agents' own choices, and the direct
mechanical consequence of transforming publicly executed actions. Separating
original choices, public reports, and executed outcomes -- rather than
treating a sharp decline in executed coordination as a single behavioural
signal -- is necessary for correctly interpreting multi-agent robustness. In
this paper, we studied how programmatic action inversion affecting both
public communication and execution shapes coordination in language-mediated
LLM multi-agent systems. Across models, honest agents' original Stag choices
decline as corrupted reports become more prevalent, while public coordination
deteriorates much more sharply. Pre-flip success does not show the same
apparent tipping pattern as public success, indicating that the executed
decline is driven substantially by deterministic inversion alongside a more
gradual behavioural response. Because pre-flip success preserves choices made
after exposure to earlier corrupted reports, it isolates the final
transformation rather than providing a no-corruption counterfactual.

Honest-agent choices are also associated with the public transcript. Under
heavier corruption, predictions based on visible public reports match honest
choices more closely than those based on hidden original actions, while prior
public Stag share is strongly associated with subsequent choices. These
patterns remain observational: the sign change in the corruption-exposure
coefficient precludes a causal-mediation interpretation, and the round-level
and speaking-position patterns do not establish a general effect of additional
interaction, although they are suggestive.

Finally, the three public-report threshold benchmarks yield similar raw match
rates. Carryover and outcome-alignment-weighted rules are slightly higher on
average, but the differences are small and model-dependent. Overall, robustness
evaluations should distinguish original choices, public actions, pre-flip
outcomes, and executed outcomes -- this separation, rather than either a
general noisy-coordination result or a claim about model rationality, is the
central methodological contribution of this work. Future work should examine open-ended
communication, longer interactions, heterogeneous groups, and adaptive
corruption strategies.

\section*{Limitations}

The experiments use homogeneous model groups, structured binary reports,
fixed speaking order, a joint transcript-and-execution intervention, and short
interaction horizons. The realised parameter grid is unbalanced, and the
focal $N=5,M=3$ setting does not contain a matched zero-corruption condition.
Pre-flip success is not a no-corruption counterfactual, and fixed order may
leave position-composition confounding. The mean-field benchmark is not an
exact sequential best response, while turn-level observations share
interaction histories.

The joint intervention also conflates two distinct effects: it simultaneously
changes what later agents observe and which action is used to determine the
realised outcome and payoff. The \textbf{b3} ablation (Appendix~\ref{app:ablations})
varies the \emph{type} of corruption applied (deterministic inversion vs.\
random relabeling) but still applies that corruption to both the transcript
and the executed action; it therefore does not decouple the communication
channel from the execution channel. Isolating the causal effect of misleading
communication alone would require a communication-only condition -- a
corrupted transcript paired with truthful execution -- which we leave for
future work.

Relatedly, because the experimental grid does not provide a fully matched
zero-corruption baseline for every model and configuration, differences in
apparent robustness to corruption across models (e.g., \texttt{Llama-3.1-8b}'s
uniformly lower benchmark match rates in \S\ref{sec:benchmark-alignment}) may partly
reflect differences in baseline strategic competence rather than differential
sensitivity to corrupted communication specifically; the present design
cannot fully separate these two explanations.

The three analytical threshold benchmarks in \S\ref{sec:methods} are also
descriptive, uncorrected reference rules rather than corruption-aware optimal
policies: as noted in Appendix~\ref{app:additional-derivations}, they do not
use the known or estimated corruption rate to adjust the observed
public-report estimate. Because the corruption mechanism here is a
deterministic inversion rather than ordinary independent random noise, simply
discounting reports by an assumed corruption rate is not necessarily the
correct or Bayes-optimal response. Deriving an optimal strategy that
explicitly models the corruption mechanism and agents' uncertainty about
corrupted reports is an important direction for future work.

The observational analyses do not identify latent beliefs, causal mediation,
truthfulness, or internally represented trust, limiting generalisation to
open-ended communication, heterogeneous teams, and adaptive corruption
strategies.

\section*{Ethical Considerations}

This study uses simulated interactions among LLM agents and does not involve
human participants or personal data. The corruption mechanism is implemented
in a controlled game environment to analyse multi-agent robustness, not to
enable real-world deception or harmful coordination. Because several evaluated
models are proprietary and may change over time, exact outputs may not remain
stable across API versions. The results concern model behaviour in this
synthetic setting and should not be generalised to human trust or social
behaviour.

\bibliography{references}

\appendix

\renewcommand{\thefigure}{A\arabic{figure}}
\renewcommand{\thetable}{A\arabic{table}}

\setcounter{figure}{0}
\setcounter{table}{0}

\section{Additional Derivations}
\label{app:additional-derivations}

\subsection{Outcome-Alignment Weight Update Details}
\label{app:trust-details}

Let \(z_j^t \in \{0,1\}\) indicate whether agent \(j\)'s round-$t$ public
report matches the final public outcome: a Stag report when public
coordination succeeds, or a Hare report when it fails. The outcome-alignment counts
are updated according to

\begin{equation}
a_j^{t+1}
=
a_j^t + z_j^t,
\qquad
b_j^{t+1}
=
b_j^t + (1-z_j^t),
\label{eq:trust_counts}
\end{equation}

and the next-round analytical weight is

\begin{equation}
\rho_j^{t+1}
=
\frac{a_j^{t+1}}
{a_j^{t+1}+b_j^{t+1}}.
\label{eq:trust_weight}
\end{equation}

These weights are computed retrospectively and should not be interpreted as
direct measurements of a speaker's truthfulness or an LLM's internal trust.

\subsection{First-Round Actions under the Threshold Benchmark}

Under incomplete information, equilibrium strategies in Bayesian games map
beliefs into actions \cite{Harsanyi1968}. In coordination environments with
strategic complementarities, this mapping often takes a threshold form in
posterior beliefs \cite{MorrisShin2003}.

In our sequential setting, the first speaker in a round does not observe any
current-round reports from other agents. Consequently, the public-report
estimate $\hat q_i^t$ is undefined for turns with no prior reports. The
agent's internal belief is also not directly observable or explicitly
reported.

Under the analytical cutoff benchmark defined above, the benchmark action
satisfies

\[
\text{Choose Stag if and only if } q_i^0 \geq q^*.
\]

If an action were generated according to this benchmark, choosing Stag would
be consistent with

\[
q_i^0 \geq q^*,
\]

whereas choosing Hare would be consistent with

\[
q_i^0 < q^*.
\]

These regions are implications of the analytical benchmark rather than
identified properties of the agent's latent belief. Inferring the belief
region directly from the action would require assuming that the agent already
follows the threshold rule being evaluated.

In addition to the reported action, agents provide a confidence score and a
short justification. While these elements are not formally incorporated into
the threshold model, they provide auxiliary behavioural information. They do
not, however, identify the value or location of an agent's latent belief.

Therefore, first-speaker actions are not used to recover latent beliefs in
the revised evaluation. Turns without prior public reports are excluded from
the common-support benchmark match-rate analysis.

\subsection{Comparative Statics of the Analytical Threshold}

The analytical threshold $q^*$ defined above varies systematically with the
structural parameters of the coordination environment. Since $q^*$ is
implicitly determined by the expected payoff inequality,

\[
q^* = f(N, M, R_S, R_F, H),
\]

changes in these parameters alter the public-report estimate required for
the benchmark to imply cooperation.

\paragraph{Effect of the Coordination Requirement ($M$).}

An increase in $M$ raises the number of Stag players required for successful
coordination. For any fixed estimate $q_i^t$, the probability
$P(K \geq M-1)$ decreases as $M$ increases. Consequently, the expected payoff
from choosing Stag declines, implying that a higher estimate is required for
the benchmark to imply cooperation. Because $M$ is discrete, this relationship
is expressed as

\[
q^*(M+1) \geq q^*(M).
\]

Thus, stricter coordination requirements weakly increase the threshold
associated with the cooperative action.

\paragraph{Effect of the Payoff Structure ($R_S, R_F, H$).}

An increase in the payoff from successful coordination $R_S$ raises the
expected return from choosing Stag for any given estimate $q_i^t$. As a
result, the threshold required for Stag decreases:

\[
\frac{\partial q^*}{\partial R_S} < 0.
\]

Conversely, increasing the safe payoff $H$ from choosing Hare raises the
threshold required for cooperation. These comparative statics follow
directly from the expected payoff condition defining $q^*$.

\paragraph{Effect of Corruption Intensity ($F$).}

The number of corrupted agents $F$ does not enter the analytical threshold
$q^*$ directly. The implementation records the untransformed-report proportion

\[
\alpha = \frac{N-F}{N},
\]

but does not use $\alpha$ to correct the observed public-report rate. The
implemented within-round estimate remains

\[
\hat q_i^t = \frac{K_i^t}{n_i^t}.
\]

Increasing $F$ can nevertheless affect coordination in two ways. First, it
changes the public reports available to later agents and may therefore be
associated with changes in their original actions. Second, deterministic
action inversion directly changes the public actions used to determine
coordination success and realised payoffs.

\medskip

Taken together, $N$, $M$, and the payoff parameters determine the analytical
threshold $q^*$, while corruption changes both the observed public history
and the executed public outcome. Structural variation therefore does not
influence outcomes exclusively through latent belief formation.

\section{Prompts and Implementation Details}
\label{app:prompts}

The initial decision call uses temperature 1.0. The system prompt is:

\begin{Verbatim}[
fontsize=\footnotesize,
breaklines=true,
breakanywhere=true
]
You are {name}, an agent in a Stag Hunt coordination game.

In each round, you must decide whether to choose STAG or HARE.
You can observe the public reports of other agents who have
already spoken.

This is an N-player game with N={num_agents} and threshold
M={stag_success_threshold}. If at least M players choose STAG,
the stag hunt succeeds.

Payoffs:
- If stag succeeds (>= M STAG):
  - STAG chooser gets {payoff_stag_success}
  - HARE chooser gets {payoff_hare_when_stag_success}
- If stag fails (< M STAG):
  - STAG chooser gets {payoff_stag_fail}
  - HARE chooser gets {payoff_hare_fail}

Your goal is to help the group coordinate on the best
equilibrium. Consider what others have reported and whether
you trust them.
\end{Verbatim}

The structured response contains \texttt{reported\_action}, a confidence
value in $[0,1]$, and a one-sentence justification. We refer to the action
returned by this initial call as the original or pre-flip action. For a
designated corrupted agent, the simulator records the initial response,
inverts the public action, and makes a second call to the same model at
temperature 1.0 to regenerate a justification consistent with the flipped
action. The original confidence is retained. The transcript entry is replaced
before the next agent receives it, and outcomes and payoffs are calculated
from the public post-flip actions. Speaking order is fixed within a run,
while corrupted-agent roles are sampled once and retained across rounds.

\section{Additional Evaluation Metrics}
\label{app:additional-metrics}

\paragraph{Honest-agent outcome alignment.}
For honest agents, ex-post outcome alignment is the proportion of turns on
which the original action matches the realised public outcome: Stag when
public coordination succeeds and Hare when it fails. We use alignment rather
than accuracy because the game provides no independent ground-truth label.

\paragraph{Confidence and calibration.}
Each agent reports confidence \(c_i^t\in[0,1]\). We group honest agents by
mean confidence into eight equal-width bins and compute Expected Calibration
Error as the weighted gap between mean confidence and mean outcome alignment.
This measures calibration relative to the realised public outcome.

\paragraph{Downstream report alignment.}
For a speaker at position \(k\), downstream report alignment is the fraction
of subsequent speakers in the same round whose public report matches the
speaker's report. Because subsequent speakers share other elements of the
public history, this measure is descriptive and does not identify causal
influence.

\paragraph{Consensus entropy.}
The public-report distribution is summarised by

\[
H^t = -\sum_{a \in \{\text{S,H}\}} p_a^t \log_2 p_a^t,
\]

where \(p_a^t\) is the fraction of agents publicly reporting action \(a\).
Entropy is zero under unanimous reporting and maximal under an even split.

\section{Additional Ablations}
\label{app:ablations}

\begin{table*}[htb]
\centering
\caption{Matched round-wise public coordination across ablations. Each row
compares paired runs at a given round. $\Delta$ denotes percentage point
differences. B3 rows use the updated redux matched estimates; their earlier
McNemar $p$-values are not reused. H2 and H3 retain the original exact
McNemar tests with Holm correction.}
\label{tab:matched-roundwise-coordination}
\begin{tabular}{llcccc}
\toprule
\textbf{Comparison} & \textbf{Round} & \textbf{Base/H1} & \textbf{Target} &
\textbf{$\Delta$ (pp)} & \textbf{$p$} \\
\midrule
\multirow{4}{*}{B3 vs. Base}
& 1 & 0.611 & 0.822 & +21.11 & -- \\
& 2 & 0.649 & 0.810 & +16.09 & -- \\
& 3 & 0.710 & 0.710 & +0.00 & -- \\
& 4 & 0.720 & 0.710 & -1.08 & -- \\
\cmidrule(lr){1-6}
\multirow{4}{*}{H2 vs. H1}
& 1 & 0.600 & 0.671 & +7.14 & 0.719 \\
& 2 & 0.705 & 0.614 & -9.09 & 0.867 \\
& 3 & 0.652 & 0.696 & +4.35 & 1.000 \\
& 4 & 0.696 & 0.696 & +0.00 & 1.000 \\
\cmidrule(lr){1-6}
\multirow{4}{*}{H3 vs. H1}
& 1 & 0.609 & 0.667 & +5.80 & 1.000 \\
& 2 & 0.705 & 0.659 & -4.55 & 1.000 \\
& 3 & 0.652 & 0.652 & +0.00 & 1.000 \\
& 4 & 0.696 & 0.739 & +4.35 & 1.000 \\
\bottomrule
\end{tabular}
\end{table*}

\paragraph{Ablation conditions.}
A randomly sampled subset of 50 base parameter points is used to evaluate
ablation conditions, keeping the base-condition run count tractable. The
subset is drawn reproducibly from the fixed seed so that comparisons across
ablations are matched on the same underlying configurations. We evaluate two
families of ablations, each isolating a distinct structural dimension of the
coordination environment. Notably, we only run ablations for a limited subset
of models due to budget constraints: \texttt{gpt-5-mini},
\texttt{gpt-5.2-2025-12-11}, and \texttt{llama-3.1-8b}.

\paragraph{Corruption type (base / b3).}
In the base condition, corrupted agents always invert their public action
deterministically: an original Stag action becomes a public Hare action and
vice versa. In ablation \textbf{b3}, the corrupted agent instead draws its
public action uniformly at random from $\{\text{Stag},\text{Hare}\}$,
independently of its original decision. The base condition is therefore
systematically anti-correlated with the original action, whereas \textbf{b3}
introduces random public-action noise. Comparing the two tests whether
deterministic anti-correlated corruption differs from random corruption.
Note that this ablation varies the \emph{type} of corruption applied within
the same joint transcript-and-execution intervention; it does not decouple
the communication channel from the execution channel (see Limitations).

\paragraph{Model heterogeneity (h1 / h2 / h3).}
The base condition \textbf{h1} assigns the same LLM to every agent, producing
a homogeneous group. Ablation \textbf{h2} assigns models round-robin across
agents by agent index from a fixed model pool, mixing models within the group.
Ablation \textbf{h3} assigns the stronger and weaker models asymmetrically
between corrupted and honest roles according to the configured assignment
policy. This tests whether asymmetric model assignment is associated with
different coordination outcomes under informational corruption.

\Cref{tab:matched-roundwise-coordination} shows the results for each ablation.
The updated B3 estimates show the clearest public-coordination differences in
rounds 1 and 2, while the later-round differences are small. This suggests
that deterministic inversion and random public-action noise differ most
clearly early in the interaction, but does not establish a general long-run
ordering. The H2 and H3 comparisons are not significant; their limited
matched support does not justify a general conclusion that model
heterogeneity improves or harms coordination.

\section{Full Results}

We report the full set of experimental results to complement the main
analysis.

\begin{figure*}[t]
    \centering
    \includegraphics[width=\linewidth]
    {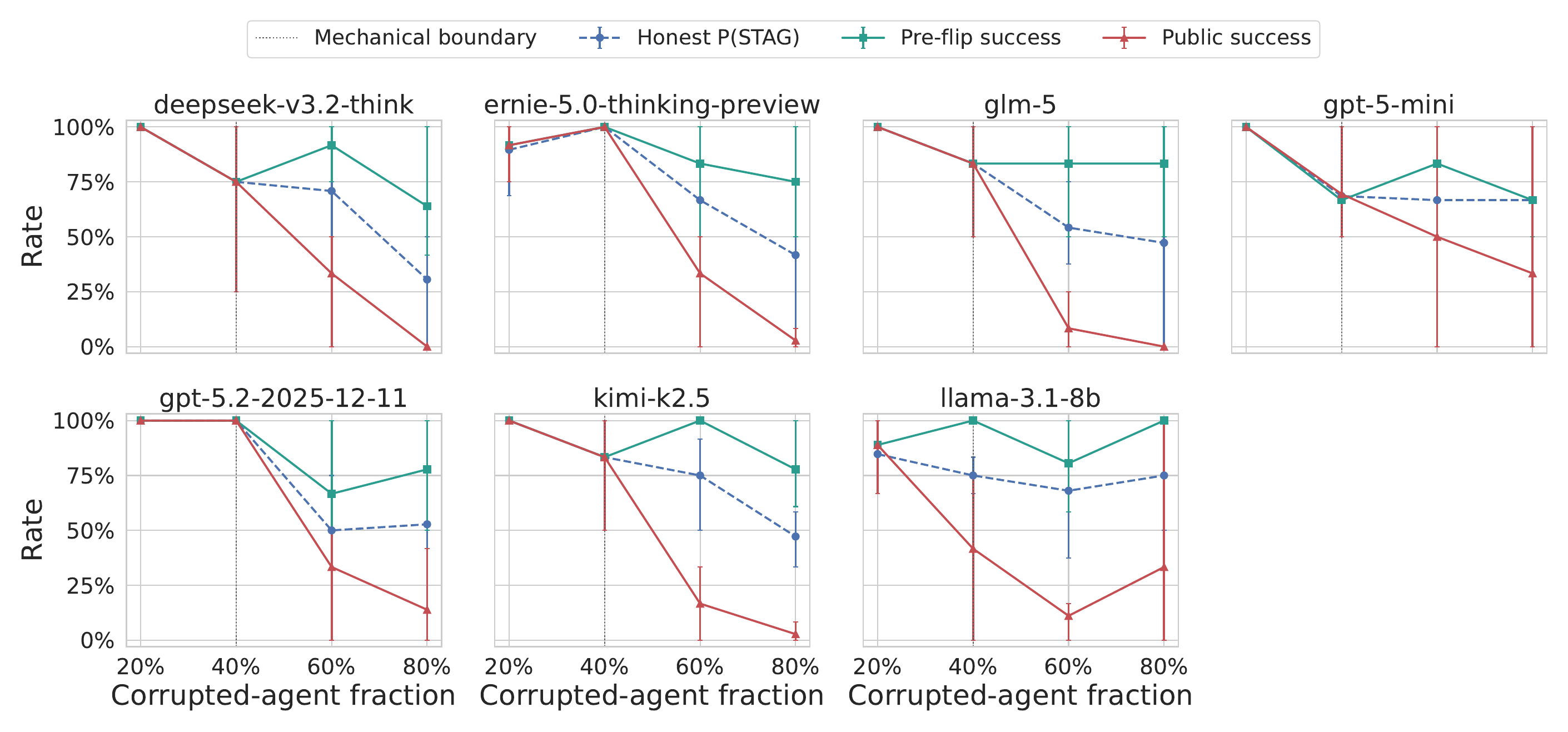}
    \caption{Mechanism decomposition by model in the focal $N=5,M=3$
    setting. Each panel separates honest original Stag rate, pre-flip
    success, and public success across corrupted-agent fractions.}
    \label{fig:mechanism-by-model}
\end{figure*}

\begin{figure*}[t]
    \centering
    \includegraphics[width=\linewidth]
    {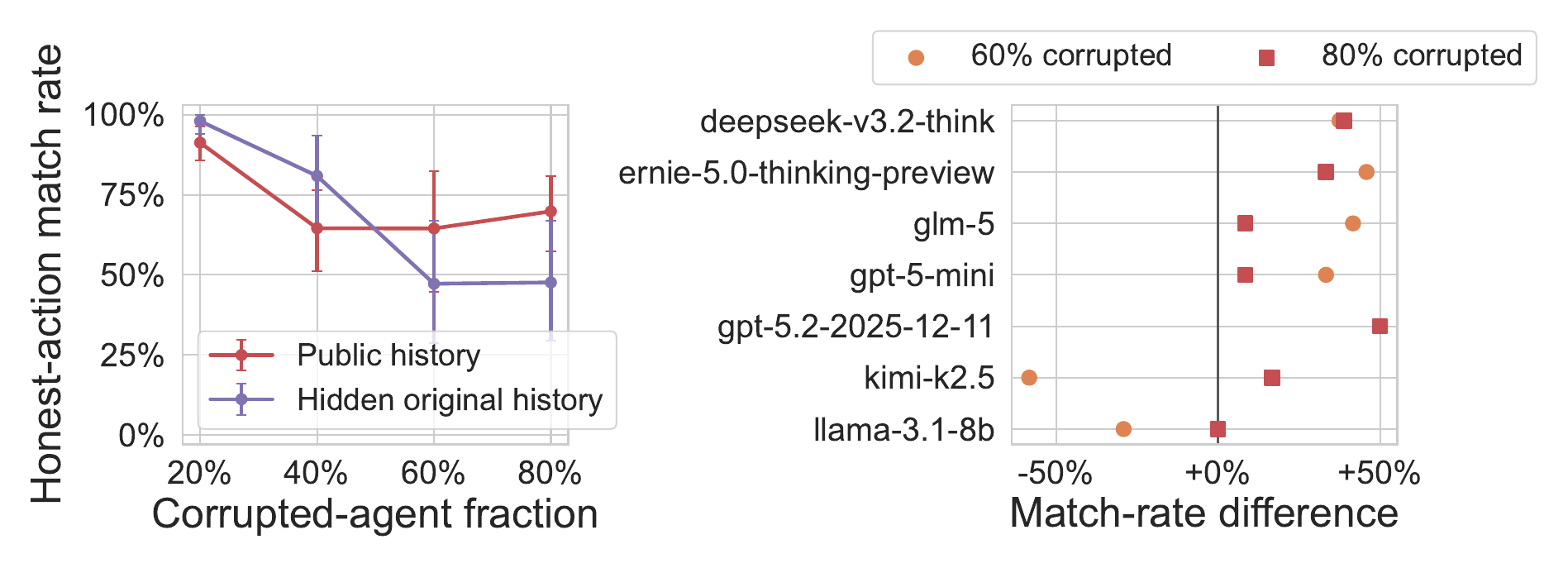}
    \caption{Match between honest original choices and threshold predictions
    constructed from the visible public history or the corresponding hidden
    original-action history in the focal $N=5,M=3$ setting.}
    \label{fig:public-vs-original-history}
\end{figure*}

\begin{figure*}[t]
    \centering
    \includegraphics[width=\linewidth]
    {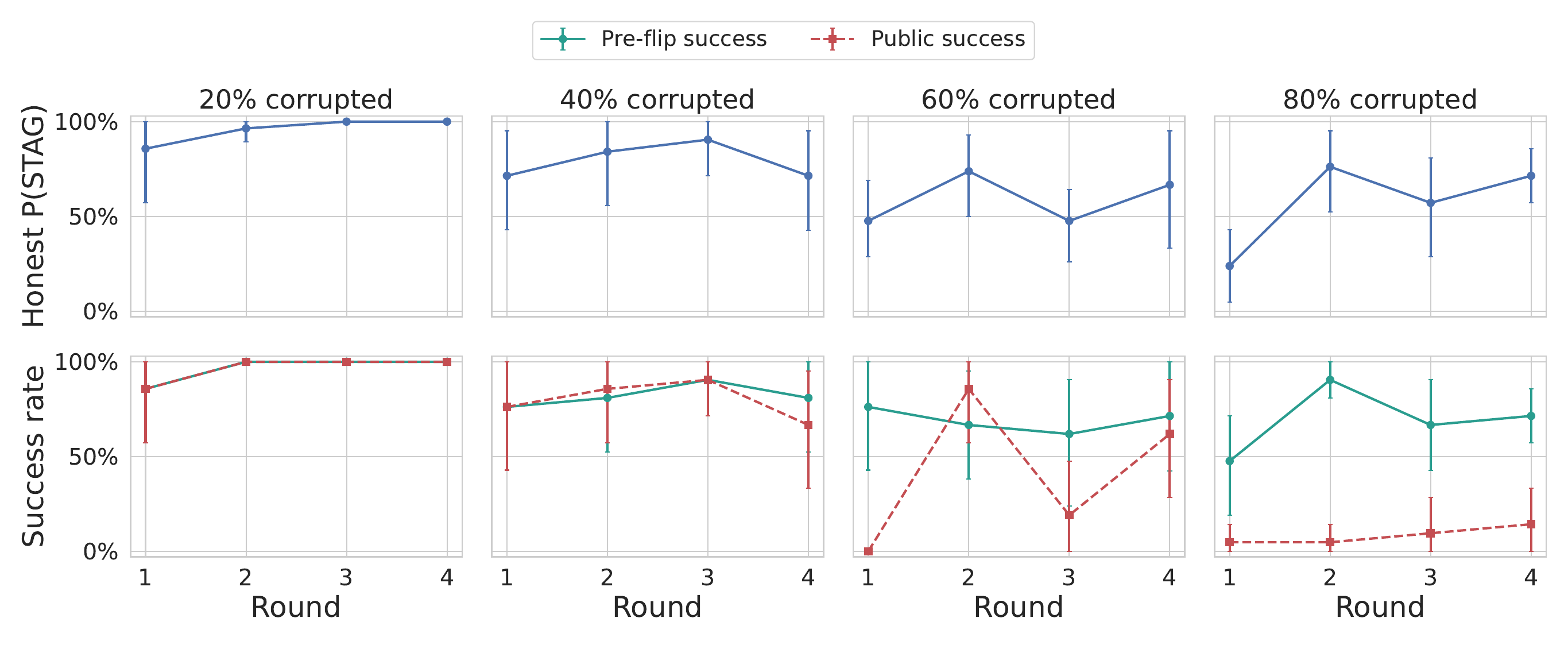}
    \caption{Honest original Stag rate, pre-flip success, and public success
    across rounds, grouped by corrupted-agent-fraction bin.}
    \label{fig:coordination-full}
\end{figure*}

\begin{figure*}[t]
    \centering
    \includegraphics[width=\linewidth]
    {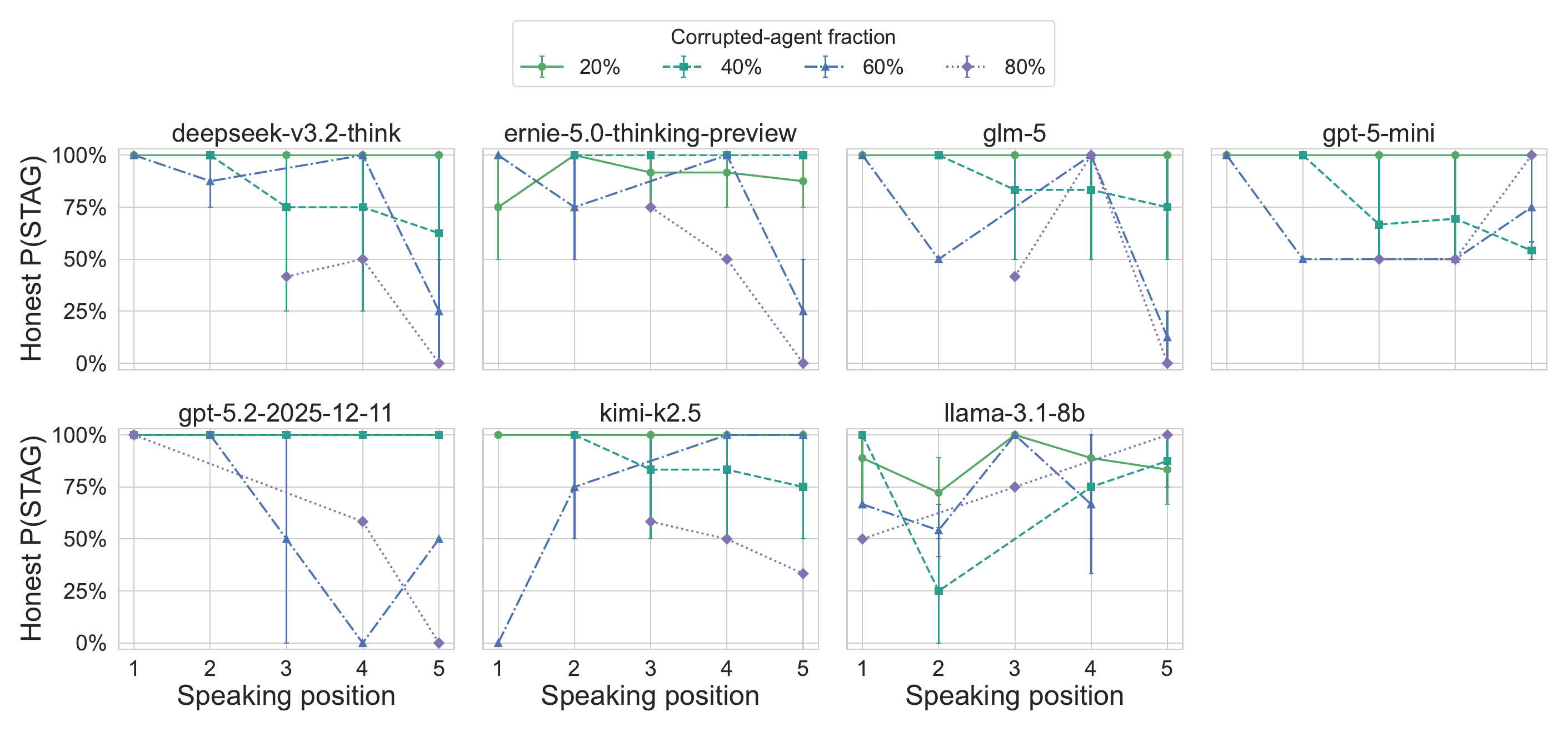}
    \caption{Honest original Stag rate by fixed speaking position and
    corrupted-agent-fraction bin. Missing position--fraction combinations
    indicate that no honest agent occupied that position in the supported
    runs.}
    \label{fig:turn-order}
\end{figure*}

\begin{figure*}[t]
    \centering
    \includegraphics[width=0.72\textwidth]
    {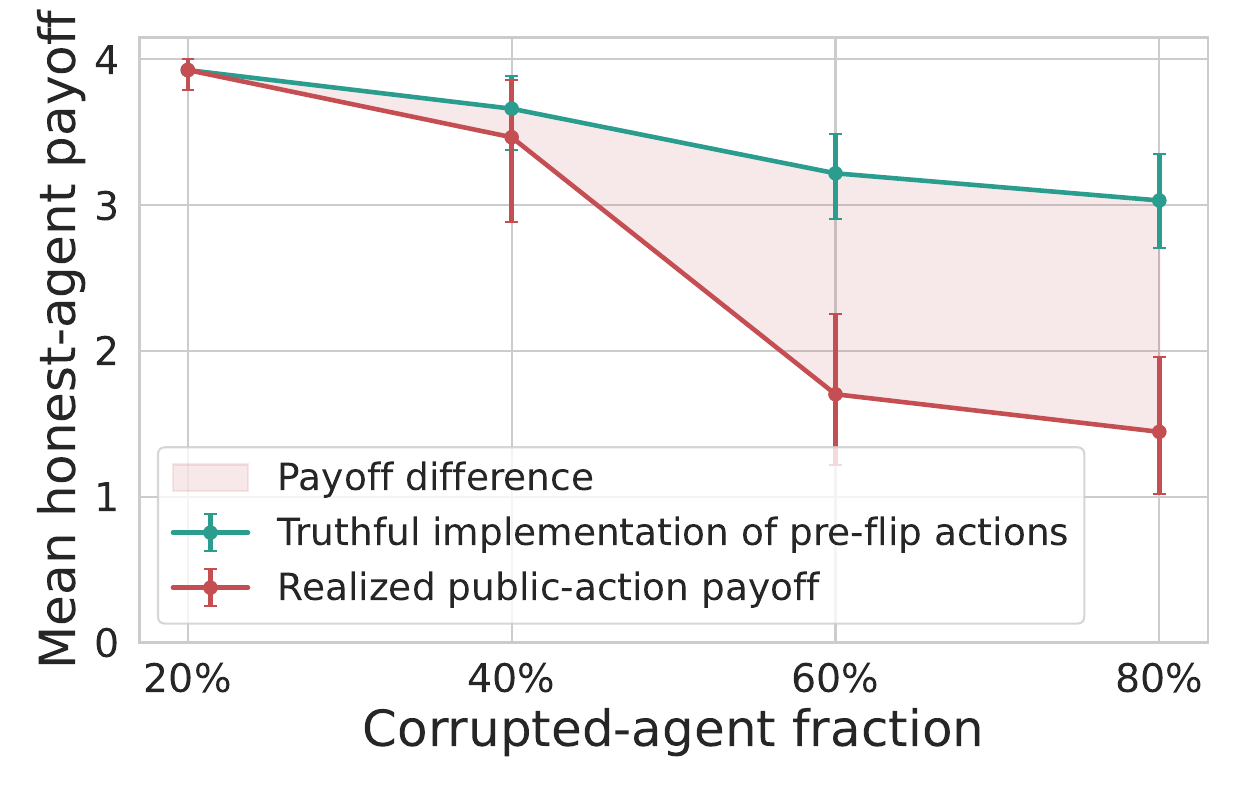}
    \caption{Honest-agent welfare decomposition in the focal $N=5,M=3$
    setting. Realised payoff is calculated from public post-flip actions and
    outcomes, while truthful-implementation payoff applies the same payoff
    rule to recorded original actions and pre-flip outcomes. Their difference
    is the mechanical payoff gap and is not a no-corruption transcript
    counterfactual.}
    \label{fig:welfare-decomposition}
\end{figure*}

\begin{figure*}[t]
    \centering
    \includegraphics[width=\linewidth]
    {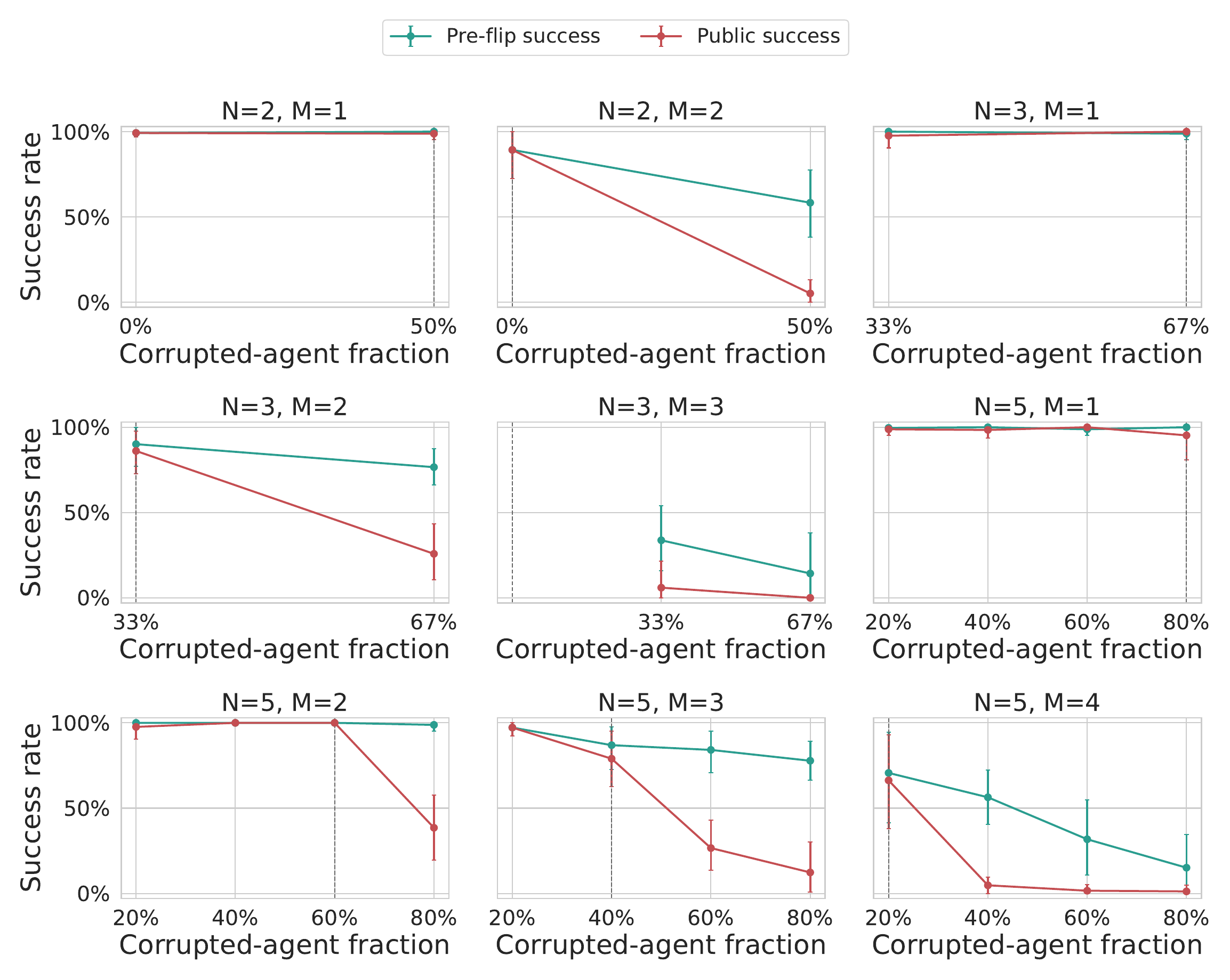}
    \caption{Pre-flip and public success rates across the supported
    combinations of $N$ and $M$. Dashed vertical lines mark the all-Stag
    inversion boundary $1-M/N$ in each panel.}
    \label{fig:coordination-full-plot}
\end{figure*}

Figure~\ref{fig:coordination-full-plot} extends the mechanism decomposition
across the supported combinations of $N$ and $M$. Pre-flip and public success
remain similar in some lower-threshold settings, whereas public success can
fall much more sharply near or beyond the all-Stag inversion boundary,
particularly under stricter coordination requirements. The persistence of
pre-flip success in several panels shows that the decline in public success
should not be interpreted as a uniform behavioural tipping point.

Table~\ref{tab:model-calibration} shows that calibration relative to the
realised public outcome varies across models, with some agents remaining
confident despite lower outcome alignment. Table~\ref{tab:overall-performance}
further shows that realised payoffs are generally lower at higher
corrupted-agent fractions. Because realised payoffs are calculated from
public post-flip actions, these differences reflect both changes in agents'
original choices and the direct effect of action inversion.
\begin{table*}[t]
    \centering
    \caption{Expected Calibration Error (ECE) relative to realised public
    outcome alignment across models.}
    \label{tab:model-calibration}
    \begin{tabular}{@{}lr@{}}
        \toprule
        \textbf{Model} & \textbf{ECE} $\downarrow$ \\
        \midrule
        ernie-5.0-thinking-preview & 0.0284 \\
        deepseek-v3.2-think        & 0.0410 \\
        gpt-5-mini                 & 0.0459 \\
        glm-5                      & 0.0572 \\
        gpt-5.2-2025-12-11         & 0.0790 \\
        kimi-k2.5                  & 0.0961 \\
        llama-3.1-8b               & 0.1186 \\
        \bottomrule
    \end{tabular}
\end{table*}

\begin{table*}[htbp]
    \centering
    \caption{Mean realised payoff across models and corrupted-agent fractions
    (mean $\pm 1.96$ SE). The 0\% rows arise only from the available
    fully honest configurations and should not be interpreted as a matched
    baseline for every group size.}
    \label{tab:overall-performance}
    \resizebox{\textwidth}{!}{%
    \begin{tabular}{lccccccc}
        \toprule
        \textbf{Corrupted-Agent Fraction}
        & \textbf{deepseek-v3.2-think}
        & \textbf{ernie-5.0-thinking-preview}
        & \textbf{glm-5}
        & \textbf{gpt-5-mini}
        & \textbf{gpt-5.2-2025-12-11}
        & \textbf{kimi-k2.5}
        & \textbf{llama-3.1-8b} \\
        \midrule
        0\%  & $4.000 \pm 0.000$ & $4.000 \pm 0.000$ & $4.000 \pm 0.000$ & $3.636 \pm 0.231$ & $4.000 \pm 0.000$ & $4.000 \pm 0.000$ & $3.556 \pm 0.317$ \\
        20\% & $3.778 \pm 0.103$ & $3.931 \pm 0.060$ & $4.000 \pm 0.000$ & $3.517 \pm 0.161$ & $4.000 \pm 0.000$ & $3.819 \pm 0.094$ & $3.057 \pm 0.187$ \\
        33\% & $3.048 \pm 0.448$ & $2.857 \pm 0.446$ & $2.857 \pm 0.520$ & $2.960 \pm 0.322$ & $2.973 \pm 0.274$ & $2.762 \pm 0.481$ & $2.439 \pm 0.379$ \\
        40\% & $3.290 \pm 0.205$ & $3.304 \pm 0.230$ & $3.261 \pm 0.235$ & $3.217 \pm 0.214$ & $3.519 \pm 0.199$ & $3.348 \pm 0.210$ & $2.725 \pm 0.298$ \\
        50\% & $2.429 \pm 0.935$ & $2.429 \pm 0.935$ & $2.429 \pm 0.935$ & $2.429 \pm 0.935$ & $3.000 \pm 0.409$ & $2.286 \pm 0.995$ & $2.625 \pm 0.856$ \\
        60\% & $2.714 \pm 0.440$ & $2.786 \pm 0.420$ & $2.643 \pm 0.413$ & $3.111 \pm 0.337$ & $3.188 \pm 0.311$ & $2.321 \pm 0.477$ & $2.120 \pm 0.299$ \\
        67\% & $2.522 \pm 0.615$ & $2.087 \pm 0.718$ & $2.609 \pm 0.575$ & $2.690 \pm 0.593$ & $2.323 \pm 0.658$ & $2.435 \pm 0.738$ & $2.074 \pm 0.609$ \\
        80\% & $2.320 \pm 0.409$ & $2.240 \pm 0.457$ & $2.240 \pm 0.414$ & $2.200 \pm 0.605$ & $2.360 \pm 0.484$ & $1.960 \pm 0.468$ & $2.150 \pm 0.452$ \\
        \bottomrule
    \end{tabular}%
    }
\end{table*}

% \begin{table*}[htbp]
%     \centering
%     \caption{Model-by-model descriptive match rates across the analytical benchmark rules.
%     The highest raw match rate for each model is bolded; no inferential
%     comparison between rules is claimed.}
%     \label{tab:belief-match-rate}
%     \begin{tabular}{lcccc}
%         \toprule
%         \textbf{Model}
%         & \textbf{Naive Aggregate}
%         & \textbf{Carryover}
%         & \textbf{Outcome-Alignment}
%         & \textbf{Best Rule} \\
%         \midrule
%         deepseek-v3.2-think        & $0.888$ & $0.908$ & $\mathbf{0.910}$ & Outcome-alignment \\
%         ernie-5.0-thinking-preview & $0.863$ & $0.873$ & $\mathbf{0.875}$ & Outcome-alignment \\
%         glm-5                      & $0.845$ & $0.865$ & $\mathbf{0.880}$ & Outcome-alignment \\
%         gpt-5-mini                 & $0.867$ & $0.886$ & $\mathbf{0.902}$ & Outcome-alignment \\
%         gpt-5.2-2025-12-11         & $0.901$ & $0.954$ & $\mathbf{0.957}$ & Outcome-alignment \\
%         kimi-k2.5                  & $\mathbf{0.848}$ & $0.843$ & $0.840$ & Naive aggregate \\
%         llama-3.1-8b               & $0.620$ & $\mathbf{0.622}$ & $0.618$ & Carryover \\
%         \bottomrule
%     \end{tabular}
% \end{table*}

\begin{table*}[htbp]
    \centering
    \caption{Model-by-model descriptive match rates across the analytical benchmark rules.
    The highest raw match rate for each model is bolded; no inferential
    comparison between rules is claimed.}
    \label{tab:belief-match-rate}

    \small
    \setlength{\tabcolsep}{4pt}

    \begin{tabularx}{\linewidth}{
        @{}
        >{\raggedright\arraybackslash}X
        >{\centering\arraybackslash}p{0.14\linewidth}
        >{\centering\arraybackslash}p{0.11\linewidth}
        >{\centering\arraybackslash}p{0.17\linewidth}
        >{\raggedright\arraybackslash}p{0.18\linewidth}
        @{}
    }
        \toprule
        \textbf{Model}
        & \textbf{Naive Aggregate}
        & \textbf{Carryover}
        & \textbf{Outcome-Alignment}
        & \textbf{Best Rule} \\
        \midrule
        deepseek-v3.2-think
            & $0.888$ & $0.908$ & $\mathbf{0.910}$ & Outcome-alignment \\
        ernie-5.0-thinking-preview
            & $0.863$ & $0.873$ & $\mathbf{0.875}$ & Outcome-alignment \\
        glm-5
            & $0.845$ & $0.865$ & $\mathbf{0.880}$ & Outcome-alignment \\
        gpt-5-mini
            & $0.867$ & $0.886$ & $\mathbf{0.902}$ & Outcome-alignment \\
        gpt-5.2-2025-12-11
            & $0.901$ & $0.954$ & $\mathbf{0.957}$ & Outcome-alignment \\
        kimi-k2.5
            & $\mathbf{0.848}$ & $0.843$ & $0.840$ & Naive aggregate \\
        llama-3.1-8b
            & $0.620$ & $\mathbf{0.622}$ & $0.618$ & Carryover \\
        \bottomrule
    \end{tabularx}
\end{table*}

\end{document}